\documentclass[11pt]{article}
\usepackage[margin=1in]{geometry}
\usepackage[T1]{fontenc}
\usepackage{lmodern,microtype,amsmath,amssymb,amsthm,booktabs,array}
\usepackage[colorlinks=true,linkcolor=blue,citecolor=blue,urlcolor=blue]{hyperref}
\usepackage{enumitem}
\usepackage{hyperref}
\usepackage{tikz}
\usetikzlibrary{arrows.meta}
\setlist{nosep}
\newcommand{\cl}{\operatorname{cl}}

\newtheorem{theorem}{Theorem}
\title{Exact Storage-Repair-Bandwidth Tradeoffs for $(6,5,5)$ and $(7,6,6)$ Exact-Repair Regenerating Codes}
\author{Chao Tian}

\begin{document}
\maketitle
\vspace{-1.5em}
\begin{abstract}
This note provides exact information-theoretic outer bounds that completely characterize the storage-repair-bandwidth tradeoff regions for the cases $(n,k,d)=(6,5,5)$ and $(7,6,6)$. The layered code proposed by Tian et al. achieves the tradeoff and is thus optimal. GPT-6 Astra assisted in the discovery of these bounds, based on a new conditioning argument and a structured search over the restricted subsets of entropy linear program constraints. The proofs are given as exact integer-valued certificates, which are difficult for direct human verification, and a Python verifier is included as part of the package, together with the certificates.
\end{abstract}

\section{Introduction}

The optimal storage-repair-bandwidth tradeoff of exact-repair regenerating codes has not been fully characterized in the literature, even though the functional repair counterpart has been resolved \cite{dimakis2010network}. Tian \cite{tian4} provided an information-theoretic characterization for the $(n,k,d)=(4,3,3)$ case, {\em i.e.,} when there are $n=4$ storage nodes, any $k=3$ of the nodes can completely recover the stored data, and any failed node can be repaired by the remaining $d=3$ nodes. This result also established a strict performance separation from its functional repair counterpart. A subsequent work \cite{tian5} resolved the case for $(n,k,d)=(5,4,4)$. Both results were obtained through an entropy linear program  (LP) \cite{yeung1997framework} that took into account the symmetry in the storage code and functional-dependence reductions. Using an iterative search approach based on heuristic inequality-subset selection, dual-supported reuse, and construction-guided equality filtering, the authors of \cite{ct2022} further characterized most of the optimal tradeoff for the case $(n,k,d)=(6,5,5)$, except for a single segment. The layered code \cite{layered}, together with space sharing, achieves the optimal tradeoffs for $(4,3,3)$ and $(5,4,4)$ (and matches the $(6,5,5)$ outer bound in \cite{ct2022} except the one segment).

Considering the restricted class of \textit{linear} exact-repair regenerating codes, Prakash and Krishnan \cite{linear} derived converse bounds to characterize the complete storage-repair-bandwidth tradeoff with $k=d=n-1$ for arbitrary $n$, showing that layered codes \cite{layered} are optimal within this class. An alternative proof of the same linear code outer bounds was given in \cite{duursma2018shortened}. Without assuming linearity, 
Mohajer and Tandon \cite{mt} established outer bounds that characterize the low-bandwidth portion of the tradeoff for
$k=d=n-1$, specifically when $\beta\le 2\alpha/k$, where $\beta$ is the repair bandwidth and $\alpha$ is the storage cost. Sasidharan et al.\ \cite{sasidharan} developed further
information-theoretic outer bounds, including an improvement of \cite{mt} when $k<d$. The linear-code converse bound in \cite{linear} holds when $k=d<n-1$, and Elyasi and Mohajer \cite{elyasi2016determinant} subsequently constructed
determinant codes achieving the linear-code converse for arbitrary $k=d\le n-1$, thereby characterizing the complete tradeoff within the linear-code class.

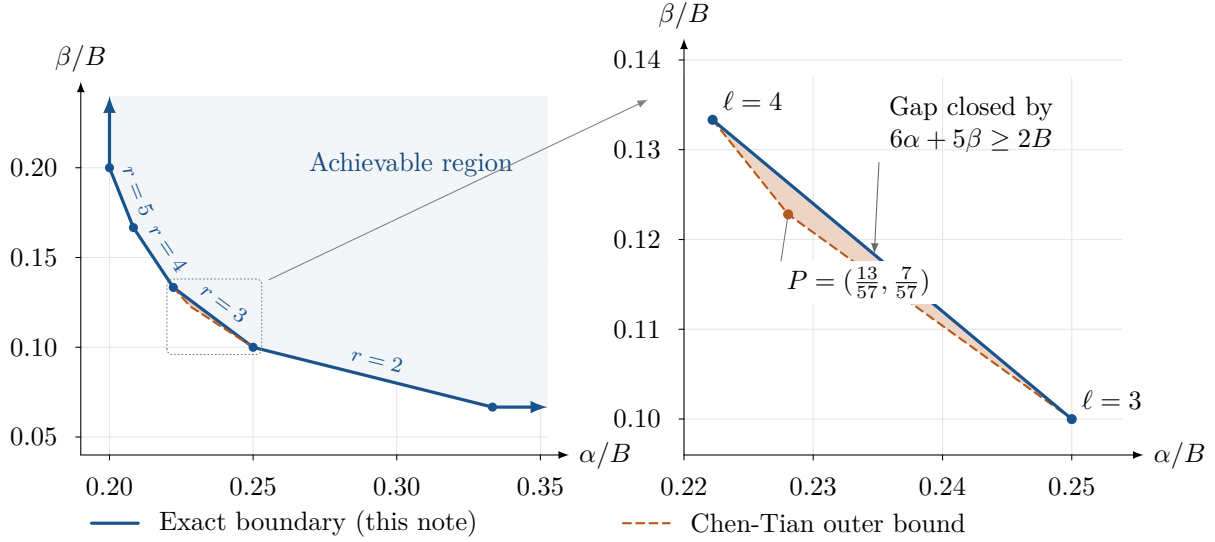
\begin{figure}[tbp]
\centering
\begingroup
\definecolor{frblue}{RGB}{24,83,142}
\definecolor{frorange}{RGB}{183,89,24}
\begin{tikzpicture}[font=\small,>=latex,line cap=round,line join=round,scale=0.95]
  \coordinate (V6) at ({40*(1/5-.19)},{25*(1/5-.04)});
  \coordinate (V5) at ({40*(5/24-.19)},{25*(1/6-.04)});
  \coordinate (V4) at ({40*(2/9-.19)},{25*(2/15-.04)});
  \coordinate (V3) at ({40*(1/4-.19)},{25*(1/10-.04)});
  \coordinate (V2) at ({40*(1/3-.19)},{25*(1/15-.04)});
  \coordinate (Pfull) at ({40*(13/57-.19)},{25*(7/57-.04)});
  \foreach \x/\lab in {.20/0.20,.25/0.25,.30/0.30,.35/0.35}{
    \draw[black!10] ({40*(\x-.19)},0)--({40*(\x-.19)},5.0);
    \draw ({40*(\x-.19)},0)--++(0,-.07) node[below=3pt] {$\lab$};
  }
  \foreach \y/\lab in {.05/0.05,.10/0.10,.15/0.15,.20/0.20}{
    \draw[black!10] (0,{25*(\y-.04)})--(6.5,{25*(\y-.04)});
    \draw (0,{25*(\y-.04)})--++(-.07,0) node[left=3pt] {$\lab$};
  }
  \fill[frblue!5] (V6)--(.4,5.0)--(6.5,5.0)--(6.5,{25*(1/15-.04)})
    --(V2)--(V3)--(V4)--(V5)--cycle;
  \node[frblue!85!black] at (4.6,4.05) {Achievable region};
  \fill[frorange!30] (V4)--(Pfull)--(V3)--cycle;
  \draw[frblue,very thick,->] (V6)--(.4,5.0);
  \draw[frblue,very thick] (V6)--node[midway,above,sloped,font=\scriptsize] {$r=5$} (V5)
    --node[midway,above,sloped,font=\scriptsize] {$r=4$} (V4)
    --node[midway,above,sloped,font=\scriptsize] {$r=3$} (V3)
    --node[midway,above,sloped,font=\scriptsize] {$r=2$} (V2);
  \draw[frblue,very thick,->] (V2)--(6.5,{25*(1/15-.04)});
  \draw[frorange,thick,densely dashed] (V4)--(Pfull)--(V3);
  \foreach \v in {V6,V5,V4,V3,V2}{\fill[frblue] (\v) circle (1.8pt);}
  \draw[->] (0,0)--(6.8,0) node[right] {$\alpha/B$};
  \draw[->] (0,0)--(0,5.2) node[above] {$\beta/B$};
  \draw[black!50,densely dotted,rounded corners=2pt]
    ({40*(.220-.19)},{25*(.096-.04)}) rectangle
    ({40*(.253-.19)},{25*(.138-.04)});
  \draw[->,black!50] (2.62,2.44)--(7.9,4.95);
  \begin{scope}[shift={(8.4,0)}]
    \coordinate (A) at ({180*(2/9-.220)},{125*(2/15-.096)});
    \coordinate (P) at ({180*(13/57-.220)},{125*(7/57-.096)});
    \coordinate (C) at ({180*(1/4-.220)},{125*(1/10-.096)});
    \foreach \x/\lab in {.22/0.22,.23/0.23,.24/0.24,.25/0.25}{
      \draw[black!10] ({180*(\x-.220)},0)--({180*(\x-.220)},5.25);
      \draw ({180*(\x-.220)},0)--++(0,-.07) node[below=3pt] {$\lab$};
    }
    \foreach \y/\lab in {.10/0.10,.11/0.11,.12/0.12,.13/0.13,.14/0.14}{
      \draw[black!10] (0,{125*(\y-.096)})--(6.1,{125*(\y-.096)});
      \draw (0,{125*(\y-.096)})--++(-.07,0) node[left=3pt] {$\lab$};
    }
    \fill[frorange!25] (A)--(P)--(C)--cycle;
    \draw[frorange,thick,densely dashed] (A)--(P)--(C);
    \draw[frblue,very thick] (A)--(C);
    \foreach \v in {A,C}{\fill[frblue] (\v) circle (2pt);}
    \fill[frorange] (P) circle (2pt);
    \node[anchor=south west] at (A) {$\ell=4$};
    \node[anchor=south west] at (C) {$\ell=3$};
    \draw[black!55] (P)--++(-.1,-.65)
      node[below right,black,fill=white,inner sep=2pt]
      {$P=(\frac{13}{57},\frac7{57})$};
    \node[align=center] (gap) at (4.0,4.6) {Gap closed by\\$6\alpha+5\beta\ge2B$};
    \draw[->,black!60] (gap.south west)--(2.65,2.8);
    \draw[->] (0,0)--(6.4,0) node[right] {$\alpha/B$};
    \draw[->] (0,0)--(0,5.8) node[above] {$\beta/B$};
  \end{scope}
  \draw[frblue,very thick] (.15,-.95)--(.8,-.95);
  \node[anchor=west] at (.95,-.95) {Exact boundary (this note)};
  \draw[frorange,thick,densely dashed] (7.55,-.95)--(8.2,-.95);
  \node[anchor=west] at (8.35,-.95) {Chen-Tian outer bound};
\end{tikzpicture}
\endgroup
\caption{Storage-repair-bandwidth region and a previous bound \cite{ct2022}. The new
 $r=3$ segment joins $(2/9,2/15)$ and $(1/4,1/10)$. The dashed segments satisfy $27\alpha+15\beta=8B$ and
 $26\alpha+25\beta=9B$. They meet at $(13/57,7/57)$, where with $6\alpha+5\beta=(113/57)B<2B$. }
\label{fig:six-node-gap}
\end{figure}

Despite the progress, the optimal tradeoff when the codes are not necessarily restricted to linear codes is still unknown. The gap between the inner bound and the outer bound in \cite{ct2022} raises the suspicion that even for $n=6$, nonlinear codes may achieve strictly better performance than linear codes. This note provides full characterizations of the tradeoff for the cases $(6,5,5)$ and $(7,6,6)$, for which the layered code \cite{layered} is again optimal. The difference between the new bound and the best known outer bound \cite{ct2022} for $(6,5,5)$ is illustrated in Fig. \ref{fig:six-node-gap}.

It should be noted that the conventional linear programming approach (e.g., the original approach used for \cite{tian4,tian5}) has no hope of finding these bounds: the two cases have at least $30$ and $42$ random variables, respectively; the corresponding LP will have on the order of $2^{30}$ and $2^{42}$ variables and more constraints, even after various reductions. In fact, even enumerating these variables to perform the reduction will be exceedingly expensive. 

The proof was obtained by GPT-6 Astra, which introduces a conditioning reduction technique and explores restricted constraint subsets in the linear programming framework, the latter of which has a flavor similar to the approach proposed in \cite{ct2022}. Exact integer certificates are provided for the converse bounds, which can be difficult to use for direct human verification. A Python verifier is therefore included with the package. The conditioning reduction bears some similarity to the shortened code technique in \cite{duursma2018shortened}, though a different conditioning set is chosen. The structural restriction of the search itself may offer some insight, even though explicit and extensible chain-style proofs have not been found at this point.

\section{Main Result}
The storage system has a total of $n$ nodes. When a node $j$ fails, the remaining nodes will send repair messages to repair the node. 
Let $S_{i\to j}$ be the repair message sent from node $i$ to node $j$. Let $E=\{S_{i\to j}:i\ne j\}$, and
\begin{equation}
 O_i=(S_{i\to j}:j\ne i),\qquad U_i=(S_{j\to i}:j\ne i),\qquad
 \label{eq:repair}
\end{equation}
from which and the repair requirement, it follows that 
\begin{equation}
 H(O_i\mid U_i)=0.
 \label{eqn:repairtoo}
\end{equation}
Let $W_i$ denote the stored content at node $i$. 

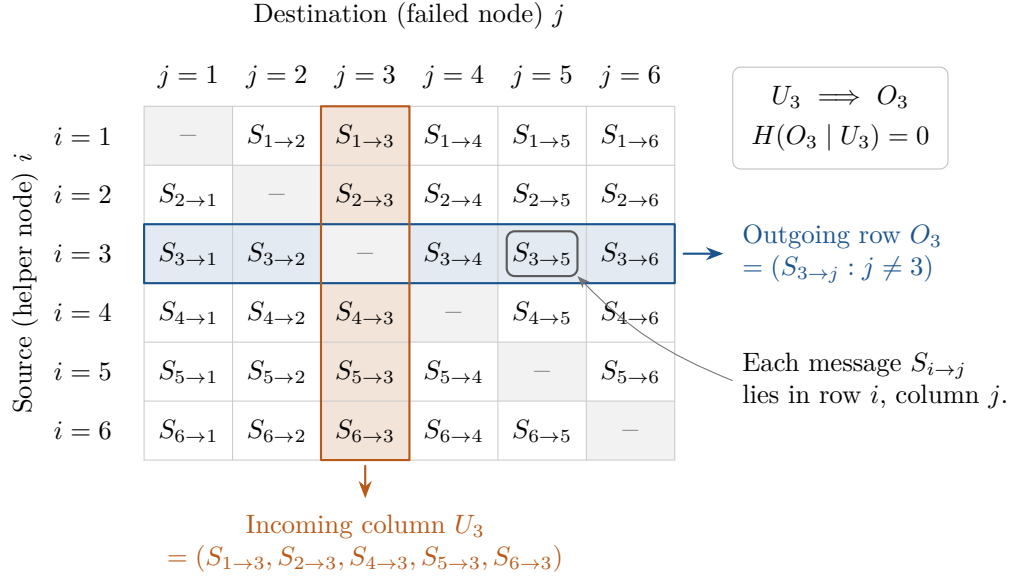
\begin{figure}[tbp]
\centering
\begingroup
\definecolor{frblue}{RGB}{24,83,142}
\definecolor{frorange}{RGB}{183,89,24}
\begin{tikzpicture}[font=\small,>=Stealth,line cap=round,line join=round,
                   x=1.17cm,y=.78cm]
  \fill[frblue!12] (.5,-3.5) rectangle (6.5,-2.5);
  \fill[frorange!16] (2.5,-6.5) rectangle (3.5,-.5);
  \foreach \i in {1,...,6}{
    \fill[black!5] ({\i-.5},{-\i-.5}) rectangle ({\i+.5},{-\i+.5});
  }
  \foreach \a in {.5,1.5,...,6.5}{
    \draw[black!20] (\a,-.5)--(\a,-6.5);
    \draw[black!20] (.5,-\a)--(6.5,-\a);
  }
  \foreach \i in {1,...,6}{
    \node[anchor=east] at (.25,-\i) {$i=\i$};
    \node at (\i,0) {$j=\i$};
    \foreach \j in {1,...,6}{
      \ifnum\i=\j
        \node[black!45] at (\j,-\i) {$-$};
      \else
        \node at (\j,-\i) {$S_{\i\to\j}$};
      \fi
    }
  }
  \draw[frblue,thick] (.5,-3.5) rectangle (6.5,-2.5);
  \draw[frorange,thick] (2.5,-6.5) rectangle (3.5,-.5);
  \node[anchor=south] at (3.5,.65) {Destination (failed node) $j$};
  \node[rotate=90] at (-.85,-3.5) {Source (helper node) $i$};
  \draw[frorange,->,thick] (3,-6.6)--(3,-7.15);
  \node[frorange,anchor=north,align=center] at (3,-7.25)
    {Incoming column $U_3$\\$=(S_{1\to3},S_{2\to3},S_{4\to3},S_{5\to3},S_{6\to3})$};
  \draw[frblue,->,thick] (6.6,-3)--(7.05,-3);
  \node[frblue,anchor=west,align=left] at (7.15,-3)
    {Outgoing row $O_3$\\$=(S_{3\to j}:j\ne3)$};
  \draw[black!65,thick,rounded corners=3pt]
    (4.6,-2.62) rectangle (5.4,-3.38);
  \node[anchor=west,align=left] (entry) at (7.15,-5.15)
    {Each message $S_{i\to j}$\\lies in row $i$, column $j$.};
  \draw[->,black!55] (entry.west) to[bend left=12] (5.4,-3.38);
  \node[draw=black!25,rounded corners=3pt,fill=white,align=center,
        inner sep=7pt,anchor=west] at (7.15,-.7)
    {$U_3\ \Longrightarrow\ O_3$\\[3pt]
     $H(O_3\mid U_3)=0$};
\end{tikzpicture}
\endgroup
\caption{Column-row structure of the repair messages, illustrated for
 $n=6$. }
\label{fig:message-column-row}
\end{figure}

For a deterministic exact-repair code, $O_i$ is a function of the stored content at node-$i$, which can be determined from $U_i$. 
All repair messages thus determine the file, and 
\begin{equation}
H(E)=B,\quad H(O_i)\leq \alpha,\quad H(S_{i\to j})\leq \beta,
\end{equation}
where $B$ is the file size, $\alpha$ is the storage cost, and $\beta$ the repair bandwidth per helper. 
The repair messages can be visualized as a matrix, given in Fig. \ref{fig:message-column-row}.

\begin{theorem}
\label{thm:main}
For $(6,5,5)$ and $(7,6,6)$ exact-repair regenerating codes, the closure of the achievable storage-repair-bandwidth region is exactly the nonnegative region specified by
\begin{align}
 (n-1)\alpha&\ge B,\label{eq:endpoints}
 \\
 nr(r-1)\alpha+n(n-1)\beta&\ge r(r+1)B,
 &&1\le r\le n-1.\label{eq:facets}
\end{align}
The converse applies without a linear-code assumption, and the optimal tradeoff is achieved by the layered code \cite{layered} and space sharing.
\end{theorem}

\begin{samepage}
The sloped boundary segments, in reduced integer form, are
\begin{center}
\begin{tabular}{c | l l}
\toprule
$r$ & $n=6$ & $n=7$\\
\midrule
2 & $2\alpha+5\beta\ge B$ & $7\alpha+21\beta\ge3B$\\
3 & $6\alpha+5\beta\ge2B$ & $7\alpha+7\beta\ge2B$\\
4 & $36\alpha+15\beta\ge10B$ & $42\alpha+21\beta\ge10B$\\
5 & $4\alpha+\beta\ge B$ (cutset) & $70\alpha+21\beta\ge15B$\\
6 & --- & $5\alpha+\beta\ge B$ (cutset)\\
\bottomrule
\end{tabular}
\end{center}
\end{samepage}
Achievability follows from the layered constructions \cite{layered} and space sharing. In the layered code, a parity group of size $\ell$ has $\ell-1$ independent symbols; placing an independent group on every $\ell$-subset gives the tradeoff point
\begin{equation}
 \left(\alpha ,\beta \right)
 =\left(\frac{\ell\cdot B}{n(\ell-1)},\frac{\ell\cdot B}{n(n-1)}\right),
 \qquad 2\le\ell\le n.
 \label{eq:vertices}
\end{equation}
The facet indexed by $r$ joins the points $\ell=r$ and $\ell=r+1$. The vertical bound $r=1$ is from a direct cut-set argument, and the horizontal bound can be obtained the same way.  The outer bounds for the segments with index $r=2$ are implied by the results given in \cite{mt}. For $n=6$, the bounds for the segments other than $r=3$ have been found previously in \cite{ct2022}.

\section{Entropy Deficits and Shortened Codes via Conditioning}
\label{sec:finite-certificates}

The quantities in \eqref{eq:facets} can be bounded by the entropies of the corresponding random variables, and the difference between the left-hand side and the right-hand side therefore represents an entropy deficit. To establish the stated inequalities, it is sufficient to show the deficit is non-negative. The proof of Theorem \ref{thm:main} requires comparing the entropy deficits of two storage systems with different numbers of storage nodes, where the smaller system can be viewed as a shortened version of the larger one via a conditioning operation. The shortened code view has been explored in \cite{duursma2018shortened}, however, the conditioning argument and conditioning sets were chosen differently from this note.

Intuitively, the argument is as follows: assume the repair messages to a single node are made available to all other nodes and repair messages; then the remaining part of the system forms a small regenerating code system, and the non-negative property of the entropy deficit for a smaller system needs to hold on this system. To make this intuition rigorous, we carry out the proof for a general class of symmetric submodular set
functions, rather than for the entropy functions of the storage codes. This stronger formulation
allows the smaller-node inequality to be applied to a contracted function.
Only after completing the induction do we specialize to the entropy function of an exact-repair code.

\subsection{Admissible Submodular Systems}

For $m\ge3$, let
\[
 E_m=\{S_{i\to j}:i,j\in\{1,\ldots,m\},\ i\ne j\},\qquad
 O_i=\{S_{i\to j}:j\ne i\},\qquad
 U_i=\{S_{j\to i}:j\ne i\}.
\]
Here the messages are formal ground-set elements. Let $\mathcal C_m$ be the
class of real-valued set functions $h:2^{E_m}\to\mathbb R$ satisfying
\begin{enumerate}
 \item normalization: $h(\varnothing)=0$;
 \item monotonicity: $h(A)\le h(B)$ whenever $A\subseteq B$;
 \item submodularity:
 $h(A)+h(B)\ge h(A\cup B)+h(A\cap B)$;
 \item node symmetry: $h(\pi A)=h(A)$ for every permutation $\pi$ of the
 $m$ node indices;
 \item repair implication: $h(O_i\mid U_i)=0$ for every $i$.
\end{enumerate}
For an arbitrary set function in this class, conditional notation means
\begin{equation}
 h(Y\mid X):=h(X\cup Y)-h(X).
 \label{eq:rank-increment}
\end{equation}
Normalization refers only to the empty set; no condition such as
$h(E_m)=1$ is imposed, and $h(E_m)=0$ is allowed. Note that such a set function does not necessarily have an entropic realization.

We say that $X$ functionally implies $Y$ if $h(Y\mid X)=0$.
Monotonicity and submodularity give
\begin{equation}
 0\le h(Y\mid B)\le h(Y\mid A),\qquad A\subseteq B.
 \label{eq:implication-monotonicity}
\end{equation}
Thus an implication remains valid after enlarging its conditioning set.
The chain identity
$h(Y\cup Z\mid X)=h(Y\mid X)+h(Z\mid X\cup Y)$ also shows
that implications compose. These facts justify every repair-closure step
below entirely within the submodular framework.

For $h\in\mathcal C_m$ and $1\le r\le m-1$, define
\begin{equation}
 D_{m,r}[h]
 =m(m-1)h(S_{1\to2})+mr(r-1)h(O_1)-r(r+1)h(E_m).
 \label{eq:deficit}
\end{equation}
Singleton braces are suppressed in expressions such as $h(S_{1\to2})$.
Symmetry makes the choice of representative row and message immaterial.
The statement to be proved is
\begin{equation}
 \mathsf P(m,r):\qquad D_{m,r}[h]\ge0
 \quad\hbox{for every }h\in\mathcal C_m,
 \label{eq:universal-induction}
\end{equation}
which would imply Theorem \ref{thm:main} when specializing the submodular function to the entropy function. 
We wish to establish this statement for $3\le m\le7$ and $1\le r\le m-1$.  
In particular, the induction hypothesis at $m-1$ concerns every member of
$\mathcal C_{m-1}$, with no assumption that it comes from a code or has a
specified total rank. 

\subsection{Incoming Message Contraction}

Let $h\in\mathcal C_m$ and assume $m\geq 4$, and consider a system with the first node removed. Write
\[
 E_{-1}=\{S_{i\to j}:2\le i,j\le m,\ i\ne j\},\qquad
 O_i^-=O_i\cap E_{-1},\qquad U_i^-=U_i\cap E_{-1}.
\]
Define a set function on the surviving messages by
\begin{equation}
 g(X)=h(X\cup U_1)-h(U_1),\qquad X\subseteq E_{-1}.
 \label{eq:poly-child}
\end{equation}
We verify that, after relabeling the surviving nodes, $g\in\mathcal C_{m-1}$.
First, $g(\varnothing)=0$, and monotonicity is inherited from $h$.
For $X,Y\subseteq E_{-1}$, submodularity of $h$ gives
\[
 g(X)+g(Y)-g(X\cup Y)-g(X\cap Y)\ge0,
\]
because $(X\cup U_1)\cap(Y\cup U_1)=(X\cap Y)\cup U_1$.
Every permutation $\pi$ of the surviving nodes $2,\ldots,m$ can extend
to a permutation of all $m$ nodes by fixing node 1. Such a permutation
only rearranges the messages in the incoming column $U_1$, so
$\pi U_1=U_1$. The symmetry of $h$ therefore gives
\[
g(\pi X)
=h\bigl(\pi(X\cup U_1)\bigr)-h(U_1)
=h(X\cup U_1)-h(U_1)
=g(X).
\]
Thus $g$ is symmetric under permutations of the surviving nodes.

It remains to check the repair implication relation. For a surviving node $i$, put
$T_i=U_i^-\cup U_1$. Since $h(O_1\mid U_1)=0$, the set $T_i$
functionally implies $S_{1\to i}$ and hence all of
$U_i=U_i^-\cup\{S_{1\to i}\}$. The parent repair implication
$h(O_i\mid U_i)=0$, together with
\eqref{eq:implication-monotonicity} and composition of implications,
therefore yields $h(O_i^-\mid T_i)=0$. Equivalently,
\begin{equation}
 g(O_i^-\mid U_i^-)
 =h(O_i^-\mid U_i^-\cup U_1)=0,\qquad 2\le i\le m.
 \label{eq:conditional-repair}
\end{equation}
This proves all the defining conditions of $\mathcal C_{m-1}$.

This smaller system must be evaluated using the submodular function $g$. Because $E_m=E_{-1}\cup U_1\cup O_1$ and
$h(O_1\mid U_1)=0$, we have
\begin{align}
 g(E_{-1})&=h(E_m)-h(U_1),
 \label{eq:child-file}\\
 g(O_2^-)&=h(O_2\mid U_1),
 \label{eq:child-row}\\
 g(S_{2\to3})&=h(S_{2\to3}\mid U_1).
 \label{eq:child-message}
\end{align}
For the row identity, the omitted message $S_{2\to1}$ already belongs to
$U_1$. Consequently, for $2\le r\le m-2$, the deficit of $g$ is
\begin{align}
 D_{m-1,r}[g]
 ={}&(m-1)(m-2)h(S_{2\to3}\mid U_1)\notag\\
 &+(m-1)r(r-1)h(O_2\mid U_1)
 -r(r+1)\bigl(h(E_m)-h(U_1)\bigr).
 \label{eq:poly-child-deficit}
\end{align}
Since $g\in\mathcal C_{m-1}$, the universal induction hypothesis
$\mathsf P(m-1,r)$ now gives
\begin{equation}
 D_{m-1,r}[g]\ge0.
 \label{eq:conditional-nonnegative}
\end{equation}
Therefore, the contracted system provides a conditional version of the entropy deficit for the larger system. 

\subsection{Starting Cases and Finite Induction Steps}

Fix $r\ge2$. The starting case is $m=r+1$. Define
\begin{eqnarray}
R=\bigcup_{i=1}^{m-2}O_i\cup\{S_{m\to m-1}\},
\qquad R'=R\cup O_{m-1}.
\end{eqnarray}
Since $U_{m-1}\subseteq R$, monotonicity and submodularity,
together with the imposed repair equality, give
\begin{eqnarray}
0\le h(O_{m-1}\mid R)
\le h(O_{m-1}\mid U_{m-1})=0.
\end{eqnarray}
Thus $h(R')=h(R)$. Similarly, $U_m\subseteq R'$, so
\begin{eqnarray}
0\le h(O_m\mid R')
\le h(O_m\mid U_m)=0.
\end{eqnarray}
Because $R'\cup O_m=E_m$, it follows that
\begin{eqnarray}
h(E_m)=h(R')=h(R)
\le \sum_{i=1}^{m-2}h(O_i)+h(S_{m\to m-1})
=(m-2)h(O_1)+h(S_{1\to2}),
\end{eqnarray}
where the inequality uses subadditivity and the last equality
uses symmetry. Consequently,
\begin{eqnarray}
D_{m,m-1}[h]
=m(m-1)\bigl((m-2)h(O_1)+h(S_{1\to2})-h(E_m)\bigr)
\ge0,
\end{eqnarray}
which is the starting case $m=r+1$.

For each pair $4\le m\le7$, $2\le r\le m-2$, the supplied certificate
gives an identity of the form
\begin{equation}
 L\bigl(pD_{m,r}[h]-qD_{m-1,r}[g]\bigr)
 =\sum_t w_t I_h(X_t;Y_t\mid R_t),
 \label{eq:poly-certificate}
\end{equation}
where $L,p,q,w_t$ are positive and
\[
 I_h(X;Y\mid R)
 :=h(X\cup R)+h(Y\cup R)-h(X\cup Y\cup R)-h(R)\ge0.
\]
The last inequality follows from monotonicity and submodularity. The
certificate identities use only repair-closure equalities and node
symmetry, which hold for every $h\in\mathcal C_m$; their validity does
not require entropic realizability. Combining
\eqref{eq:poly-certificate} with \eqref{eq:conditional-nonnegative} gives
\begin{equation}
 D_{m,r}[h]
 =\frac qp D_{m-1,r}[g]
   +\frac1{Lp}\sum_t w_t I_h(X_t;Y_t\mid R_t)\ge0.
 \label{eq:induction-step}
\end{equation}

For completeness, the endpoint arguments also apply to every member of
$\mathcal C_m$. Any $m-1$ outgoing rows imply the last row and hence
$E_m$. The lower triangular set $\{S_{i\to j}:i>j\}$ implies
$O_1,O_2,\ldots,O_m$ successively and hence also $E_m$.
Subadditivity and symmetry give
\begin{equation}
 h(E_m)\le(m-1)h(O_1),\qquad
 h(E_m)\le\binom m2 h(S_{1\to2}).
 \label{eq:entropy-endpoints}
\end{equation}
The second inequality is $\mathsf P(m,1)$, completing the stated finite
family of inequalities.

\subsection{Entropy Function as Specialization}

Now consider a deterministic exact-repair code.
Let $h_0(A)=H(A)$ be the joint entropy of the repair messages in $A$.
The functional relations $H(O_i\mid W_i)=0$ and $H(W_i\mid U_i)=0$
imply $h_0(O_i\mid U_i)=0$. Thus $h_0$ satisfies normalization,
monotonicity, submodularity, and the repair implications, although it need
not be symmetric. As noted in Section~2, determinism and file recovery
give
\[
 h_0(E_n)=B,\qquad h_0(O_i)\le\alpha,\qquad
 h_0(S_{i\to j})\le\beta.
\]
Define the symmetrized set function
\begin{equation}
 \bar h(A)=\frac1{n!}\sum_{\pi\in\mathfrak S_n}h_0(\pi A).
 \label{eq:entropy-symmetrization}
\end{equation}
The defining axioms and repair equalities are preserved by relabeling
and averaging, so $\bar h\in\mathcal C_n$. Furthermore,
\[
 \bar h(E_n)=B,\qquad \bar h(O_1)\le\alpha,\qquad
 \bar h(S_{1\to2})\le\beta.
\]
Applying $\mathsf P(n,r)$ yields
\begin{align}
 r(r+1)B
 &\le nr(r-1)\bar h(O_1)+n(n-1)\bar h(S_{1\to2})\notag\\
 &\le nr(r-1)\alpha+n(n-1)\beta,
 \qquad 1\le r\le n-1.
 \label{eq:deficit-to-rate}
\end{align}
Induction from $m=r+1$ will prove $\mathsf P(m,r)$ for $m=4,5,6,7$, when the certificates in the form of (\ref{eq:poly-certificate}) can be provided. The ten supplied certificates cover exactly the pairs needed; see Table \ref{tab:counts}. 

\begin{table}
\caption{Relevant counts for the certificates\label{tab:counts}}
\begin{center}
\begin{tabular}{c c r r r}
\toprule
$n$ & $r$ & $p$ ($q=1$) & Positive CMI terms & Maximum residual depth\\
\midrule
4&2&3&10&2\\
5&2&3&86&3\\
5&3&6&93&3\\
6&2&3&250&4\\
6&3&6&203&4\\
6&4&10&42&4\\
7&2&3&117&5\\
7&3&6&1144&5\\
7&4&20&501&5\\
7&5&15&213&5\\
\bottomrule
\end{tabular}
\end{center}
\end{table}
The smaller cases are included so that no unprovided child certificate is
needed. The displayed term counts are certificate sizes, not minimum proof
lengths. The largest integer scale is
$1282833116168492207230$, which exceeds the signed 64-bit
integer range. The verifier therefore uses Python's built-in
arbitrary-precision integers to perform all coefficient
calculations exactly.

\section{Restricted Search and Certificates}

\subsection{Reduced LP and Restricted Search}

The reduced LP formulation utilized the same function implication relation and symmetry relation used in \cite{tian4}. More precisely, for a message set $R$, repeatedly add the full row $O_i$ whenever the full incoming column $U_i$ is present. Denote the resulting set by $\cl(R)$. Equation~\eqref{eqn:repairtoo} and repeated application of the repair relation give $H(R)=H(\cl(R))$. Sets related by a node permutation have equal entropy under symmetry. 

Only CMIs tight on both complete parity layers of sizes $r$ and $r+1$ are retained, which is one of the heuristics used in \cite{ct2022}. The target on the left-hand side in \eqref{eq:poly-certificate} vanishes on both layers. Thus any positive decomposition of this target can use only terms that vanish there. This filter is logically necessary for the chosen target; it does not impose layered entropy values on an arbitrary code. For a block of size $\ell$, $t$ distinct exposed symbols have entropy $\min(t,\ell-1)$, which makes the filter exactly checkable.

For closed $R$, let
\begin{equation}
 K(R)=\{i:O_i\subseteq R\},\qquad
 P(R)=R\setminus\bigcup_{i\in K(R)}O_i,\qquad
 \delta(R)=|P(R)|.
 \label{eq:depth}
\end{equation}
The successful dictionaries used conditioning sets with $\delta(R)\le n-2$ for $n=4,5,6,7$. Equivalently, complete outgoing rows are free to include, and only the remaining messages count toward the depth budget. The residual messages may be disconnected and may contain directed cycles. This is an edge-count restriction on a closed set, not a path-length bound. It applies to the conditioning set; the other three entropy items in the decomposed conditional mutual information inequalities are included when generated, with no separate depth requirement.

For the certificates in the form of (\ref{eq:poly-certificate}), the generated inequalities have $X$ and $Y$ each equal to a single repair message or a complete outgoing row. Whole-row arguments package multiple elementary Shannon inequalities without requiring their intermediate coordinates, which differs from the approach taken in \cite{ct2022} and may partly explain why their characterization was not complete. Combining the conditional mutual information inequalities, the entropy terms in the expanded expressions can cancel out, resulting in a proof.

The search targets \eqref{eq:poly-certificate}, rather than a direct rate inequality; in contrast, direct rate inequalities were targeted in \cite{ct2022}, which may require more conditional mutual information terms as a certificate and thus more difficult to find. The extra conditional coordinates connect the parent to a smaller valid repair system. Any finite positive $p/q$ suffices for the induction. For example, $(n,r)=(7,4)$ was certified with $p/q=20$; the attempted value 10 was not found in the tested dictionaries, though a larger dictionary may allow a bound with a lower ratio.

The accompanying verifier audits these restrictions on every retained term. All ten certificates satisfy $\delta\le n-2$, the stated argument grammar, and all terms are tight on both adjacent layers. The $(7,3)$ and $(7,4)$ certificates contain respectively 164 and 51 terms whose residual conditioning graph has a directed cycle. Therefore, these particular certificates do not belong to an acyclic-residual dictionary. 

\subsection{Certificate Files}

The directory contains ten unchanged JSON certificates, \texttt{manifest.json}, and a readable verifier \texttt{verify.py}. From the extracted directory, run
\begin{verbatim}
python verify.py
python verify.py --json verification_results.json
\end{verbatim}
The first command verifies every identity, each closure and permutation witness, positivity, the documented search restrictions, and completeness of the finite induction. It also checks the certificate hashes. It does not require an LP solver, NumPy, network access, or floating-point arithmetic. JSON labels start from 0, instead of 1. Ordered pairs $(i,j)$ with $i\ne j$ are assigned bits in lexicographic order as in Fig. \ref{fig:message-column-row}. 

The exact integer certificates, Python verifier, and accompanying
materials are available through the
\href{https://ct2641.github.io/ExactRegCode67/}{companion webpage}
and the
\href{https://github.com/ct2641/ExactRegCode67}{GitHub repository}.

\section{Conclusion}
This note documents the proof arguments and the certificates for the converse bounds of the $(6,5,5)$ and $(7,6,6)$ exact-repair regenerating codes. With these results, it becomes more plausible to conjecture that: 1) the layered code with space sharing is optimal for $(n,k,d)$ when $k=d=n-1$ and general $n$, and 2) linear codes can achieve the complete storage-repair-bandwidth tradeoff for regenerating codes. The shortened code approach and the conditioning technique can be extended in various ways, and potentially lead to proofs for more general cases. The author did attempt to produce a general proof of such a nature with AI assistance, but so far the effort has been unsuccessful. An attempt to extend the results to $(8,7,7)$ with GPT-6.1 Sol (and GPT-6 Astra) was also not successful. 

\section*{Disclosure of AI Usage}

GPT-6 Astra proposed the conditional shortened codes for the induction and conducted the restricted search, with the references \cite{tian4,tian5,ct2022,sasidharan} and the suggestion to consider induction-based proofs included in the prompt. After obtaining the initial result, the author prompted GPT-6 to find the submodularity-based argument as a rigorous proof. GPT-6 also produced the initial draft of the note, the Python verifier, and the certificates. The author verified the reasoning steps, revised the document, identified the connections to past results in the literature, and reviewed and tested the Python script. The author takes full responsibility for the correctness of the results. 

\bibliographystyle{ieeetr}
\bibliography{finite_region_references}

\end{document}